\documentclass[10pt]{iopartmod}
\usepackage{amssymb}
\usepackage{amsmath}
\usepackage{makeidx}
\usepackage[dvips]{graphicx}

\input{tcilatex}

\begin{document}

\title{Absorption suppression in photonic crystals}
\author{A. Figotin and I. Vitebskiy}

\begin{abstract}
We study electromagnetic properties of periodic composite structures, such
as photonic crystals, involving lossy components. We show that in many cases
a properly designed periodic structure can dramatically suppress the losses
associated with the absorptive component, while preserving or even enhancing
its useful functionality. As an example, we consider magnetic photonic
crystals, in which the lossy magnetic component provides nonreciprocal
Faraday rotation. We show that the electromagnetic losses in the composite
structure can be reduced by up to two orders of magnitude, compared to those
of the uniform magnetic sample made of the same lossy magnetic material.
Importantly, the dramatic absorption reduction is not a resonance effect and
occurs over a broad frequency range covering a significant portion of the
respective photonic frequency band.
\end{abstract}

\maketitle

\section{Introduction}

Magnetic materials play a crucial role in microwave technology and optics.
They are absolutely essential in numerous nonreciprocal devices such as
isolators, circulators, phase shifters, etc. They can also provide
tunability, miniaturization, better impedance matching, and other important
features. A major obstacle for broader applications of magnetic materials is
tied up with the issue of absorption. Many magnetic materials with otherwise
perfect physical characteristics have been rejected because of strong losses
at frequency range of interest. In this paper we explore the idea of
composite magnetic structures having desired physical properties associated
with magnetism but, at the same time, significantly suppressing the effects
of absorption. In other words, we want to take advantage of the useful
characteristics of a particular magnetic material, while drastically
reducing its contribution to the energy dissipation.

Magnetic photonic crystals are spatially periodic composite structures with
one of the components being a magnetic material, such as a ferromagnet or a
ferrite. Extensive information on the subject and numerous references can be
found in a recent review article \cite{MPC Inoue06}. Similarly to other
photonic crystals, magnetic photonic crystals display strong spatial
dispersion, resulting in appearance of electromagnetic band-gap structure.
But in addition, magnetic photonic crystals can provide tunability and
better impedance matching than regular non-magnetic photonic crystals \cite%
{Lyubch}.

In comparison to uniform magnetic materials, magnetic photonic crystals can
display much stronger nonreciprocal properties, such as magnetic Faraday
rotation \cite{MPC Inoue06,Lyubch,MPC
Inoue99,Levy07,Levy06,PRE01,PRB03,Grishin07,Grishin04}. Strong nonreciprocal
effects are essential in isolators, circulators, and other microwave and
optical devices. The possibility of appreciable enhancement of Faraday
rotation is particularly important at infrared and optical frequencies,
where the non-reciprocal effects in uniform magnetic materials are very
weak. In addition, the use of periodic structures instead of uniform
magnetic materials can dramatically reduce the size of non-reciprocal and
other microwave and optical devices.

The subject of our investigation is another important aspect of
electrodynamics of periodic composite structures. Namely, we show that
properly design periodic array can dramatically reduce the losses associated
with individual constitutive components. In the particular case of magnetic
photonic crystals, the broadband suppression of losses can be achieved in a
combination with the enhancement of nonreciprocal properties, such as
Faraday rotation, linear magnetoelectric response, etc. The possibility of
the reduction of losses is related to the fact that in most cases the
absorption and the useful functionality of the particular magnetic material
are related to different components of its permittivity and/or permeability
tensors $\hat{\varepsilon}$ and $\hat{\mu}$. Specifically, the absorption is
determined by the anti-Hermitian parts $\hat{\varepsilon}^{\prime \prime }$
and $\hat{\mu}^{\prime \prime }$ the permittivity and permeability tensors%
\begin{equation}
\hat{\varepsilon}^{\prime \prime }=-\frac{i}{2}\left( \hat{\varepsilon}-\hat{%
\varepsilon}^{\dag }\right) ,\ \hat{\mu}^{\prime \prime }=-\frac{i}{2}\left( 
\hat{\mu}-\hat{\mu}^{\dag }\right) ,  \label{abs}
\end{equation}%
where $\dag $ denotes Hermitian conjugate. By contrast, the nonreciprocal
circular birefringence responsible for the Faraday rotation is usually
determined by the Hermitian skew-symmetric part of the respective tensors%
\begin{equation}
\hat{\varepsilon}_{a}=\frac{1}{2}\left( \hat{\varepsilon}-\hat{\varepsilon}%
^{T}\right) ,\ \hat{\mu}_{a}=\frac{1}{2}\left( \hat{\mu}-\hat{\mu}%
^{T}\right) ,  \label{skews}
\end{equation}%
where the subscript $T$ denotes matrix transposition. The relations (\ref%
{abs}) and (\ref{skews}) suggest that the rate of energy absorption by the
lossy material can be functionally different from its useful functionality
(nonreciprocal circular birefringence in our case). Such a difference allows
us to adjust the physical and geometric characteristics of the periodic
structure so that the electromagnetic field distribution inside the photonic
crystal suppresses the energy dissipation by the lossy magnetic component,
while preserving or even enhancing its useful functionality. Moreover, in
some cases, a sufficiently strong absorption can affect the electromagnetic
field distribution in such a way that it suppresses its own contribution to
the total rate of absorption of the composite material. In the latter
situation, the stronger absorption coefficient of the lossy component is,
the less it contributes to the total rate of absorption of the composite
structure.

The way to address the problem of absorption suppression in a periodic
composite structure essentially depends on the following three factors.

\begin{enumerate}
\item[1.] The useful functionality of the lossy material. In our example, it
will be the nonreciprocal circular birefringence producing the Faraday
rotation.

\item The dominant physical mechanism of absorption. For instance, energy
dissipation caused by electric conductivity requires a different approach,
compared to the situation where the losses are associated with the dynamics
of magnetic domains, or some other physical mechanisms. In each individual
case, the structure of the anti-Hermitian part (\ref{abs}) of the
permittivity and/or permeability tensors can be different, and so can be the
optimal configuration of the composite material.

\item The frequency range of interest. The same periodic array can
significantly reduce losses at some frequencies, while enhancing losses at
different frequencies. In other words, the same periodic structure can be
either effective or counterproductive, depending on the frequency range and
the dominant physical mechanism of electromagnetic energy dissipation.
\end{enumerate}

The possibility of a significant absorption reduction is not limited to
magnetic composites. Similar approach can be applied to other heterogeneous
structures with lossy components. A key requirement is that the useful
functionality of the lossy material should be functionally different from
its contribution to the energy dissipation. In such a case, the periodic
structure can be engineered so that at frequency range of interest, the
electromagnetic field distribution inside the composite medium suppresses
the losses, while preserving the useful functionality of the lossy
component. This can always be achieved if the components of the tensors $%
\hat{\varepsilon}$ and $\hat{\mu}$ related to the useful functionality of
the lossy material are different from those dominant in the anti-Hermitian
tensors $\hat{\varepsilon}^{\prime \prime }$ and $\hat{\mu}^{\prime \prime }$%
. In any event, in order to suppress the energy dissipation caused by lossy
component of the periodic structure we have to take into account the
physical nature of absorption, the useful functionality of the lossy
material, and the frequency range of interest.

In the next section we produce a specific numerical example of a periodic
structure that includes a lossy magnetic material. The amount of Faraday
rotation produced by this periodic structure is comparable to that of a
uniform slab made of the same lossy magnetic material. But the absorption
rate of the composite structure is 20 to 100 times lower than that of the
uniform magnetic slab. Importantly, such a dramatic absorption reduction is
achieved in a broad frequency range covering almost an entire photonic
frequency band. Another practical advantage is that in some cases the size
of the composite structure can be much smaller compared to that of the
uniform magnetic sample with similar characteristics. Note, though, that a
significant size reduction usually comes at the expense of the bandwidth.

\section{Absorption suppression in a periodic layered structure}

\begin{figure}[tbph]
\scalebox{0.8}{\includegraphics[viewport=0 0 500 350,clip]{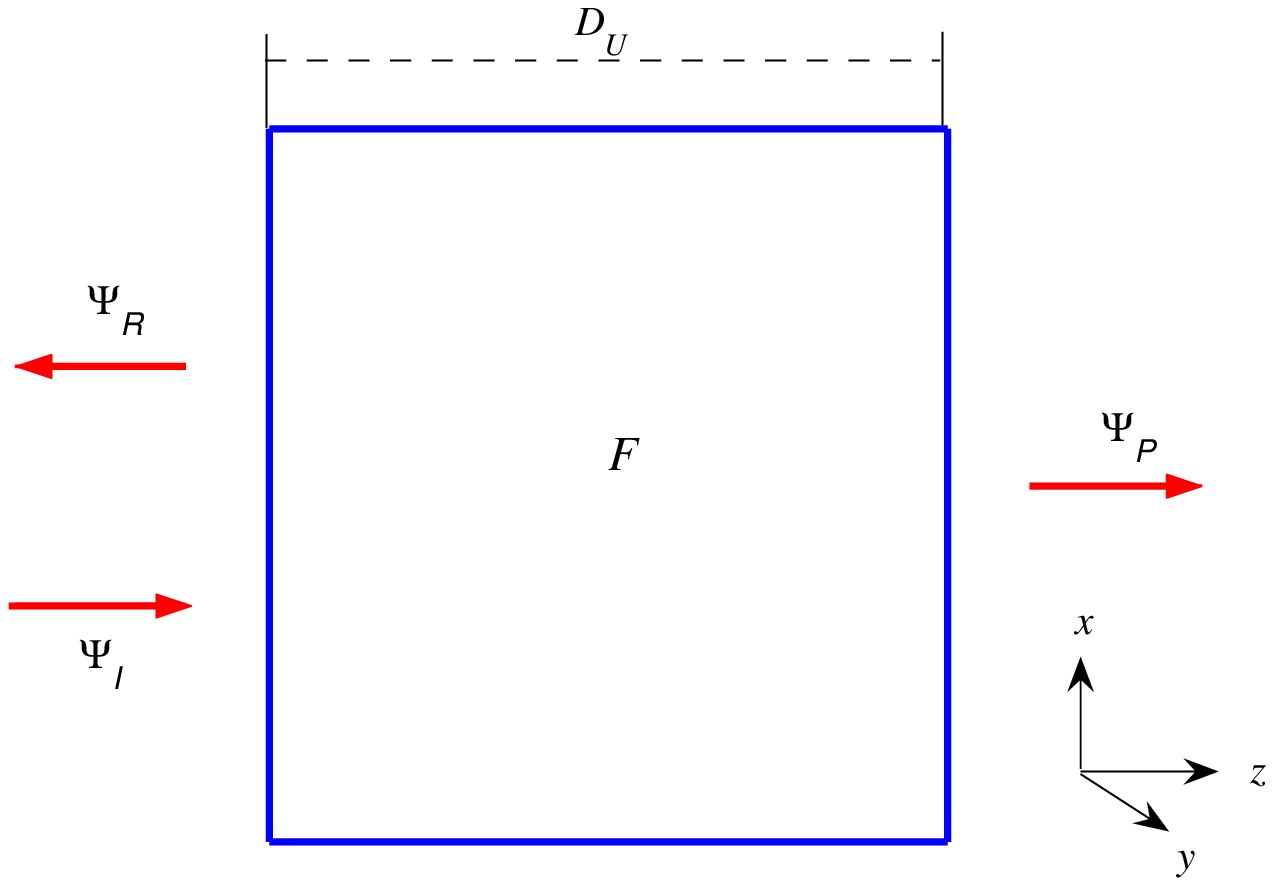}}
\caption{(Color online) Electromagnetic wave propagation through a uniform
lossy magnetic slab. The direction of magnetization is normal to the slab.
The incident wave $\Psi _{I}$ is linearly polarized with $E\parallel x$. Due
to the nonreciprocal circular birefringence/dichroism of the magnetic
material, the reflected wave $\Psi _{R}$ and the transmitted wave $\Psi _{P}$
are both elliptically polarized.}
\label{FSnU}
\end{figure}

\begin{figure}[tbph]
\scalebox{0.8}{\includegraphics[viewport=0 0 500 350,clip]{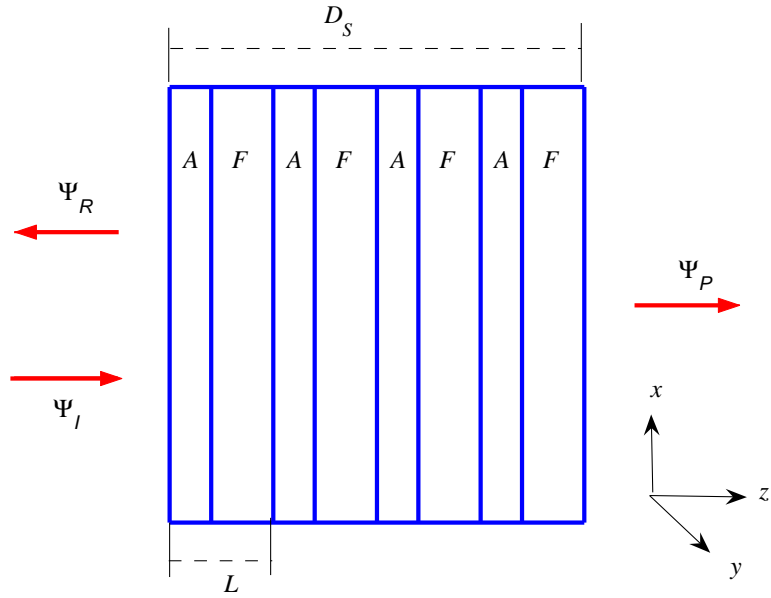}}
\caption{(Color online) The same as in Fig.\protect\ref{FSnU}, but instead
of the uniform magnetic slab we have a periodic layered structure composed
of alternate magnetic ($F$) and dielectric ($A$) layers. The $F$ layers are
made of the same lossy magnetic material as the uniform slab in Fig. \protect
\ref{FSnU}. $L$ is the unit cell length.}
\label{FSn8}
\end{figure}

Consider a monochromatic plane wave normally incident on a uniform magnetic
slab, as shown in Fig. \ref{FSnU}. Suppose that the useful functionality of
the slab is the nonreciprocal Faraday rotation. This is the case with almost
all nonreciprocal microwave and optical devices, such as the isolators,
circulators, etc. The statement of the problem is as follows. On the one
hand, we have a plane-parallel uniform magnetic slab characterized by
certain Faraday rotation and absorption. This uniform slab is shown in Fig. %
\ref{FSnU}. On the other hand, we have a stack of layers made of the same
lossy magnetic material alternating with some other layers. Such a stack is
shown in Fig. \ref{FSn8}. We expect the properly designed periodic stack in
Fig. \ref{FSn8} to be superior to the uniform magnetic slab in Fig. \ref%
{FSnU}. The superiority can be defined by the following set of requirements.

\begin{enumerate}
\item[1.1] The stack produces similar or larger Faraday rotation, as
compared to that of the uniform magnetic slab. This means that the stack and
the uniform slab have comparable useful functionality.

\item[1.2] The stack has much lower absorption than the uniform magnetic
slab. This requirement reflects our prime objective.

\item[1.3] The stack dimensions do not exceed those of the uniform magnetic
slab.

\item[1.4] The stack displays all the above properties within a reasonably
broad frequency range.
\end{enumerate}

The above set of requirements corresponds to a broadband absorption
suppression. Alternatively, we can impose a slightly different set of
requirements.

\begin{enumerate}
\item[2.1] The stack produces similar or larger Faraday rotation in
comparison to the uniform magnetic slab.

\item[2.2] The stack has much lower absorption than the uniform magnetic
slab.

\item[2.3] The stack has much smaller dimensions than the uniform magnetic
slab.
\end{enumerate}

Notice, that the requirement 2.3 of much smaller dimensions comes at the
expense of the bandwidth (the requirement 1.4).

In this section we provide numerical examples proving the effectiveness of
the photonic crystal approach to absorption suppression. In all cases we
assume that the permittivity and permeability tensors of the magnetic
material have the following standard form%
\begin{equation}
\hat{\varepsilon}_{F}=\left[ 
\begin{array}{ccc}
\varepsilon _{1}+i\gamma _{e} & i\alpha & 0 \\ 
-i\alpha & \varepsilon _{1}+i\gamma _{e} & 0 \\ 
0 & 0 & \varepsilon _{3}%
\end{array}%
\right] ,~~\hat{\mu}_{F}=\left[ 
\begin{array}{ccc}
\mu _{1}+i\gamma _{m} & i\beta & 0 \\ 
-i\beta & \mu _{1}+i\gamma _{m} & 0 \\ 
0 & 0 & \mu _{3}%
\end{array}%
\right] ,  \label{eps_F, mu_F}
\end{equation}%
corresponding to either uniaxial or isotropic magnetic material placed in a
bias magnetic field parallel to the $z$ axis \cite{LLEM,Gurev}. The
quantities $\alpha $ and $\beta $ in (\ref{eps_F, mu_F}) are responsible for
nonreciprocal circular birefringence \cite{LLEM}. If the direction of
magnetization is changed for the opposite, the parameters $\alpha $ and $%
\beta $ will also change sign and so will the sense of Faraday rotation.
Usually, at microwave frequencies, $\left\vert \beta \right\vert \gg
\left\vert \alpha \right\vert $, while at optical frequencies, $\left\vert
\beta \right\vert \ll \left\vert \alpha \right\vert $. The positive
parameters $\gamma _{e}$ and $\gamma _{m}$ in (\ref{eps_F, mu_F}) are
responsible for absorption. We can assume, for example, that the dominant
physical mechanism of absorption is the electric conductivity $\sigma $,
which is often the case at microwave frequencies. In a transverse
electromagnetic wave, the effect of electric conductivity reduces to the
following anti-Hermitian contribution to the electric permittivity tensor $%
\hat{\varepsilon}_{F}$ \cite{LLEM}%
\begin{equation}
i\gamma _{e}=4\pi \sigma i/\omega .  \label{gamma_e}
\end{equation}%
Different absorption mechanisms could also result in the dominance of the
parameter $\gamma _{m}$ in (\ref{eps_F, mu_F}), rather than $\gamma _{e}$.
In our numerical examples we only consider the case of $\gamma _{e}\gg
\gamma _{m}$.

The uniform magnetic slab in Fig. \ref{FSnU} has the thickness $D_{U}$ and
the permittivity and permeability tensors (\ref{eps_F, mu_F}). In Fig. \ref%
{FSn8} we show a setting similar to that of Fig. \ref{FSnU}, but the uniform
magnetic slab is now replaced with the periodic layered structure composed
of alternating magnetic and dielectric layers. The magnetic $F$ layers are
made of the same lossy magnetic material as the uniform slab in Fig. \ref%
{FSnU}. The respective permittivity and permeability tensors are given in (%
\ref{eps_F, mu_F}). The non-magnetic $A$ layers are made of lossless
isotropic dielectric material with the permittivity and permeability tensors%
\begin{equation}
\hat{\varepsilon}_{A}=\left[ 
\begin{array}{ccc}
\varepsilon _{0} & 0 & 0 \\ 
0 & \varepsilon _{0} & 0 \\ 
0 & 0 & \varepsilon _{0}%
\end{array}%
\right] ,~~\hat{\mu}_{A}=\left[ 
\begin{array}{ccc}
1 & 0 & 0 \\ 
0 & 1 & 0 \\ 
0 & 0 & 1%
\end{array}%
\right] .  \label{eps_A, mu_A}
\end{equation}%
The role of the $A$ layers is to create the proper electromagnetic field
distribution, which would suppress the energy absorption in the lossy $F$
layers, while preserving or even enhancing the non-reciprocal effects
associated with magnetism. The total number of the double layers $L$ in the
periodic stack is $N$. The thicknesses $d_{A}$ and $d_{F}$ of the $A$ and $F$
layers satisfy the relation $d_{A}+d_{F}=L$. The total thickness $D_{S}$ of
the periodic stack is the product $NL$. We assume that in all cases the
thickness $D_{S}$ of the periodic stack in Fig. \ref{FSn8} does not exceed
the thickness $D_{U}$ of the uniform magnetic slab in Fig. \ref{FSnU}.

The key to the possibility of absorption suppression in the composite
structure is that the nonreciprocal effects on the one hand, and the energy
dissipation on the other hand, are determined by different components of the
material tensors (\ref{eps_F, mu_F}). This allows to design the periodic
array so that at the frequency range of interest, the electromagnetic field
distribution inside the composite structure will favor the non-reciprocal
components of the permittivity/permeability tensors responsible for the
desired Faraday rotation, while not engaging with the anti-Hermitian
components responsible for absorption.

Computations of electromagnetic field distribution inside the stratified and
uniform media, as well as the amplitude and polarization of the transmitted
and reflected waves are based on the time-harmonic Maxwell equations%
\begin{equation}
\nabla \times \vec{E}\left( z\right) =i\frac{\omega }{c}\hat{\mu}\left(
z\right) \vec{H}\left( z\right) ,\;\nabla \times \vec{H}\left( z\right) =-i%
\frac{\omega }{c}\hat{\varepsilon}\left( z\right) \vec{E}\left( z\right) ,
\label{THME}
\end{equation}%
with $z$ - dependent permittivity and permeability tensors defined in (\ref%
{eps_F, mu_F}) and (\ref{eps_A, mu_A}). The electric conductivity $\sigma $
is included in the definition (\ref{eps_F, mu_F}) of the respective
permittivity tensor $\hat{\varepsilon}_{F}$, as specified in (\ref{gamma_e}%
). In all cases, the incident wave $\Psi _{I}$ propagates along the $z$
direction normal to the layers and has linear polarization with $\vec{E}%
_{I}\parallel x$. Due to the nonreciprocal circular birefringence of the
magnetic material, the transmitted and reflected waves $\Psi _{P}$ and $\Psi
_{R}$ will be elliptically polarized with the ellipse axes being at an angle
with the $x$ direction. The magnitude of nonreciprocal effects can be
characterized by the $y$ component $\left( \vec{E}_{P}\right) _{y}$ of the
transmitted wave $\Psi _{P}$. Indeed, in the absence of magnetism, the
parameters $\alpha $ and $\beta $ in (\ref{eps_F, mu_F}) vanish and the
transmitted wave will be linearly polarized with $\vec{E}_{P}\parallel x$.
This is true both in the case of the uniform magnetic slab in Fig. \ref{FSnU}
and in the case of the layered structure in Fig. \ref{FSn8}.

The transmission and reflection coefficients of the slab (either uniform, or
layered) are defined as follows%
\begin{equation}
t=\frac{S_{P}}{S_{I}},\ r=-\frac{S_{R}}{S_{I}},  \label{t,r}
\end{equation}%
where $S_{I}$, $S_{P}$, and $S_{R}$ are the Poynting vectors of the
incident, transmitted, and reflected waves, respectively. The slab
absorption is 
\begin{equation}
a=1-t-r.  \label{a}
\end{equation}%
If the incident wave polarization is linear, the coefficients $t$, $r$, and $%
a$ are independent of the orientation of vector $\vec{E}_{I}$ in the $x-y$
plane. Assuming that the incident wave $\Psi _{I}$ is linearly polarized
with $\vec{E}_{I}\parallel x$, the magnitude of the nonreciprocal effect can
be characterized by the ratio%
\begin{equation}
\rho =\frac{\left( E_{P}\right) _{y}}{\left( E_{I}\right) _{x}},\text{ \
where \ }\left\vert \rho \right\vert <1.  \label{rho}
\end{equation}

Generally, the transmitted wave polarization in Figs. \ref{FSnU} and \ref%
{FSn8} is elliptical, rather than linear. Therefore, the quantity $\rho $ in
(\ref{rho}) is not literally the sine of the Faraday rotation angle. The
nonreciprocal effects in the scattering problem of Figs. \ref{FSnU} and \ref%
{FSn8} are more complicated than a simple Faraday rotation. Let us elaborate
on this point. The electromagnetic eigenmodes of the uniform magnetic slab
in Fig. \ref{FSnU} and the eigenmodes of the layered structure in Fig. \ref%
{FSn8} are all circularly polarized. This implies that if the polarization
of the incident wave is circular, the transmitted and reflected waves will
also be circularly polarized, both in the case of a uniform slab and in the
case of a layered stack, with or without absorption. On the other hand, due
to the nonreciprocal (magnetic) effects, the transmission/reflection
coefficients for the right-hand circular polarization are different from
those for the left-hand circular polarization. This is true regardless of
the presence or absence of absorption. Consider now \ a linearly polarized
incident wave. It can be viewed as a superposition of two circularly
polarized waves with equal amplitudes. Sine the transmission/reflection
coefficients for the right-hand and left-hand circular polarizations are
different, the transmitted and reflected waves will be elliptically
polarized. Such an ellipticity develops both in the case of a uniform slab
and in the case of a layered stack, with or without absorption. Note,
though, that at optical frequencies, the dominant contribution to the
ellipticity of transmitted wave is usually associated with absorption, which
is largely responsible for circular dichroism. Without absorption, the
ellipticity of the wave transmitted through magnetic the slab in Fig. \ref%
{FSnU} would be negligible. This is not the case, though, at microwave
frequencies, where the ellipticity of transmitted and reflected waves can be
significant even in the absence of absorption.

To avoid confusion, note that a linear polarized wave propagating in a
uniform, lossless, unbounded, magnetic medium (\ref{eps_F, mu_F}) will not
develop any ellipticity. Instead, it will display a pure Faraday rotation.
But the slab boundaries in Fig. \ref{FSnU} and the layer interfaces in Fig. %
\ref{FSn8} will produce some ellipticity even in the case of lossless
magnetic material. The absorption provides an additional contribution to the
ellipticity of transmitted and reflected waves. The latter contribution is
referred to as circular dichroism. The following numerical examples
illustrate some of the above statements.

For simplicity, in further consideration we will often refer to the quantity 
$\rho $ in (\ref{rho}) as the amount of (nonreciprocal) Faraday rotation,
although, due to the ellipticity, it is not exactly the sine of the Faraday
rotation angle.

In all plots, the frequency $\omega $ and the Bloch wave number $k$ are
expressed in dimensionless units of $cL^{-1}$ and $L^{-1}$, respectively. In
our computations we use a transfer matrix approach identical to that
described in Ref. \cite{PRE01,PRB03}.

\subsection{Broadband absorption suppression}

We start with the following set of numerical values%
\begin{equation}
\begin{array}{c}
d_{A}=0.8L,\ d_{F}=0.2L,\ N=8,\ D_{S}=8L,\ D_{U}=10L, \\ 
\ \varepsilon _{1}=2.89,\ \mu _{1}=32.49,\ \varepsilon _{0}=26.01, \\ 
\alpha =0,\ \beta =4.0,\text{\ \ }\gamma _{e}=0.1,\text{\ }\gamma _{m}=0.%
\end{array}%
\   \label{set 1}
\end{equation}%
The value $\gamma _{e}=0.1$ corresponds to a relatively strong absorption
which can be associated with the electric conductivity of the magnetic
material. According to (\ref{set 1}), the thickness $D_{S}$ of the periodic
stack is somewhat smaller than the thickness $D_{U}$ of the uniform magnetic
slab. In addition, that the combined thickness $Nd_{F}$ of all magnetic
layers in the stack is 6 times smaller than the thickness $D_{U}$ of the
uniform magnetic slab. The latter implies that not only the periodic stack
in Fig. \ref{FSn8} is thinner than the uniform magnetic slab in Fig. \ref%
{FSnU}, but the actual amount of magnetic material used in the composite
structure is just a one sixth of that used in the uniform slab.

\begin{figure}[tbph]
\scalebox{0.8}{\includegraphics[viewport=-50 0 400 350,clip]{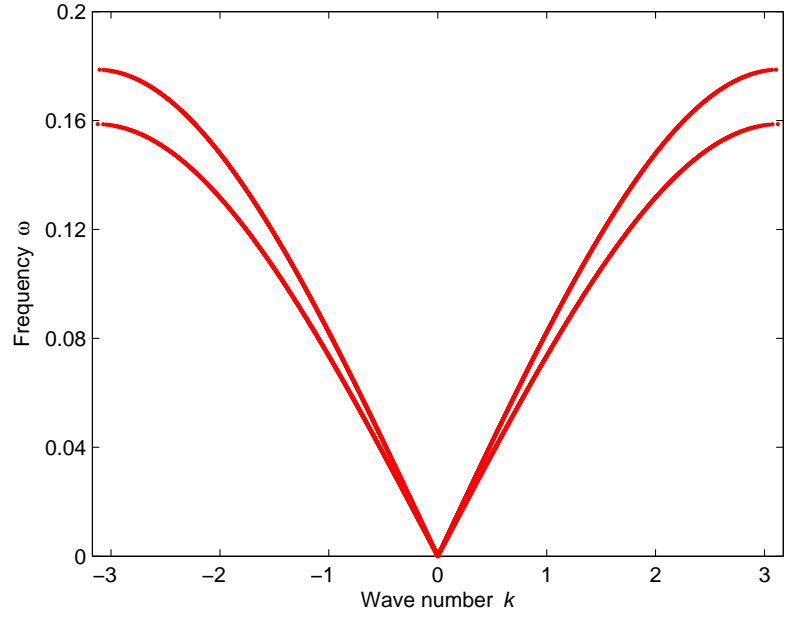}}
\caption{(Color online) The lowest band of the Bloch $k-\protect\omega $
diagram of the periodic layered structure in Fig. \protect\ref{FSn8} with
physical parameters specified in (\protect\ref{set 1}). The split of the
spectral branches is due to the circular birefringence in the magnetic
layers.}
\label{DR_a08_b4}
\end{figure}

\begin{figure}[tbph]
\scalebox{0.8}{\includegraphics[viewport=-30 0 500
400,clip]{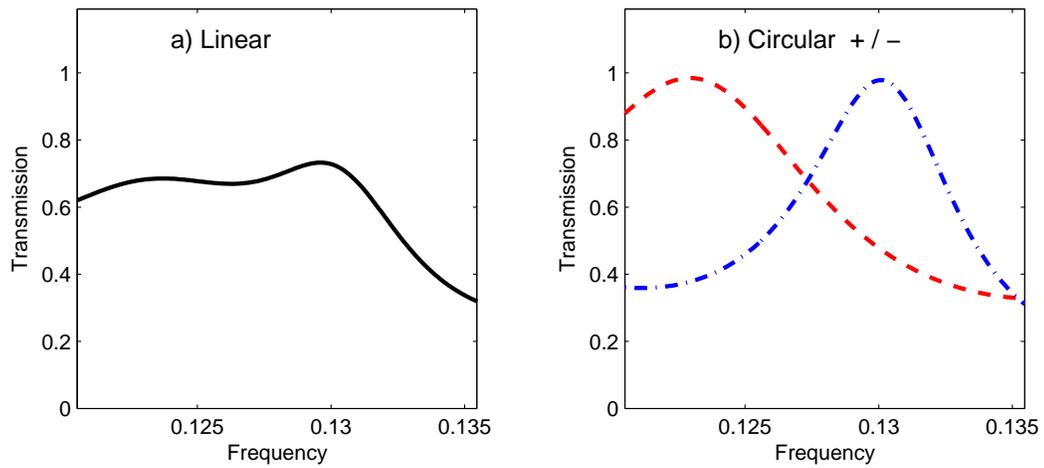}}
\caption{(Color online) Transmission $t$ versus frequency $\protect\omega $
of the periodic magnetic stack in Fig. \protect\ref{FSn8} with the physical
characteristics specified in (\protect\ref{set 1}): (a) the incident wave
polarization is linear. (b) the incident wave polarization is circular. Two
transmission maxima on the right correspond to two Fabry-Perot resonances
for each of the two senses of circular polarization. At the intersection
frequencies of the two curves, the transmitted wave polarization is linear,
while at all other frequencies, it is elliptical.}
\label{tf_a08_b4_LC}
\end{figure}

In Fig. \ref{DR_a08_b4} we present a fragment of the $k-\omega $ diagram of
the periodic array of layers described in (\ref{set 1}); only the lowest
photonic frequency band is shown. The split of the two spectral branches is
due to the strong circular birefringence $\beta =4$ in (\ref{set 1}). In
Fig. \ref{tf_a08_b4_LC} we show the transmission dispersion of the periodic
magnetic stack composed of 8 unit cells $L$. The strong polarization
dependence of the stack transmission is also caused by the large value of
the nonreciprocal parameter $\beta $ in (\ref{set 1}).

\begin{figure}[tbph]
\scalebox{0.8}{\includegraphics[viewport=-30 0 500 500,clip]{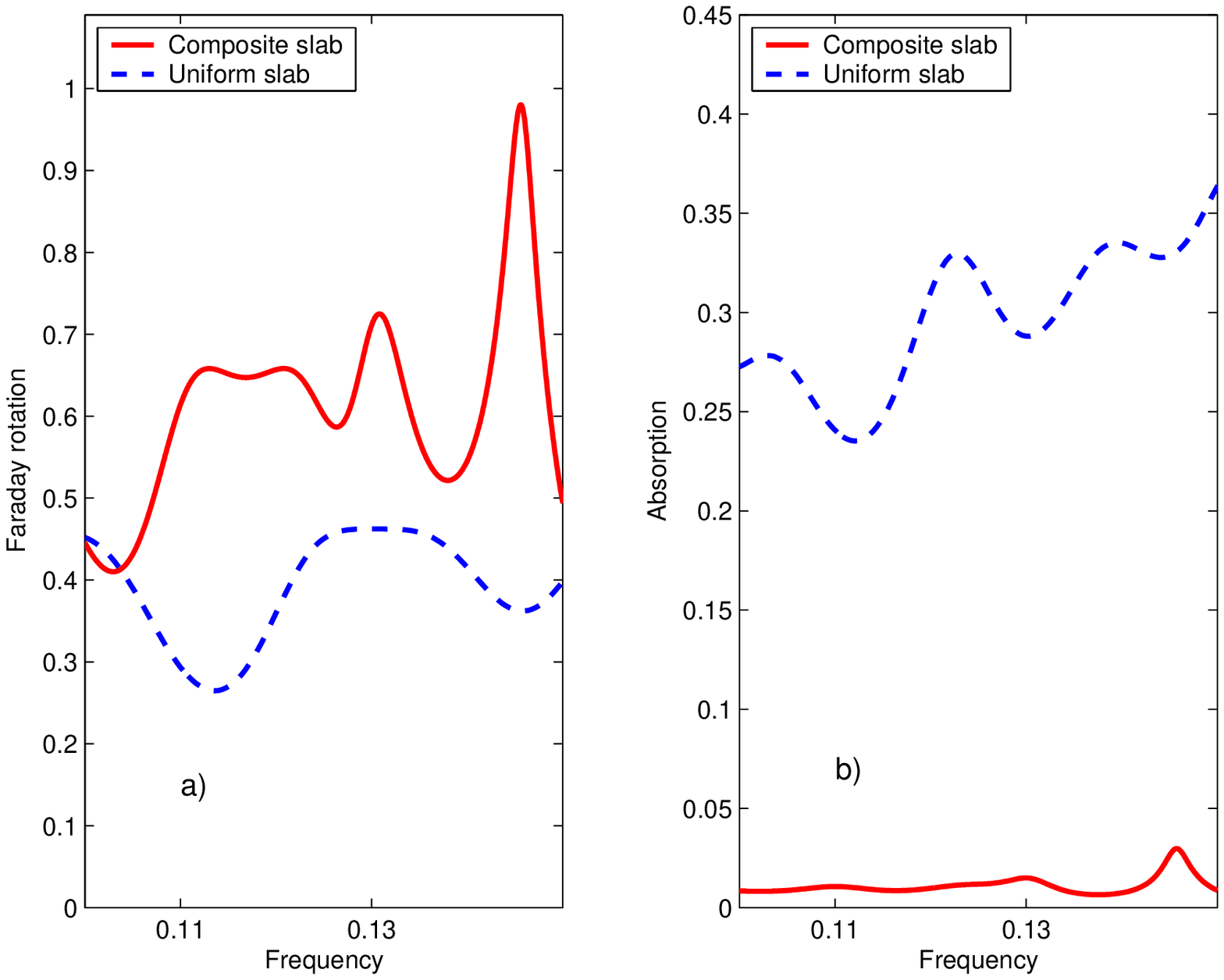}}
\caption{(Color online) Comparative transmission characteristics of the
periodic magnetic stack versus the uniform magnetic slab. The physical
parameters of both uniform and non-uniform samples are specified in (\protect
\ref{set 1}): (a) Faraday rotation $\left\vert \protect\rho \right\vert $
versus frequency $\protect\omega $ of the uniform magnetic slab (the dashed
curve) and the periodic layered structure (the solid curve). (b) The
respective absorption $a$ versus frequency $\protect\omega $. The graphs
show that the composite structure produces similar Faraday rotation as the
uniform magnetic slab, while greatly reducing the absorption in a relatively
broad frequency range. The frequency range shown coincides with that in Fig. 
\protect\ref{tf_a08_b4_LC}.}
\label{af_pt_b4}
\end{figure}

Let us now compare the performance of the uniform magnetic slab in Fig. \ref%
{FSnU} and the periodic stack in Fig. \ref{FSn8}. Fig. \ref{af_pt_b4}(a)
shows that the magnitude $\left\vert \rho \right\vert $ of the Faraday
rotation produced by the periodic stack is comparable to that of the uniform
slab. In fact, the periodic stack produces even stronger nonreciprocal
effect. At the same time, Fig. \ref{af_pt_b4}(b) shows that within the same
broad frequency range, the energy dissipation in the layered structure is 20
to 50 times lower, compared to the uniform slab. Hence, the composite
structure does indeed dramatically reduce the losses while enhancing the
useful functionality of the magnetic material and reducing the overall
dimensions. In other words, by all accounts, the layered structure is by far
superior to the uniform magnetic slab.

The numerical parameters (\ref{set 1}) used in our computations are
hypothetical, although realistic. We did not try to optimize the
configuration of the periodic layered structure, or to see if the periodic
arrays with two or three dimensional periodicity can produce even better
results. Our goal here is to prove that even a simple periodic array can
dramatically improve the situation with losses. A key is the proper
configuration of the composite structure, which, in turn, essentially
depends on the physical nature of absorption and the frequency range of
interest. Also, we would like to emphasize that the dramatic absorption
reduction seen in Fig. \ref{af_pt_b4} is not a resonance effect. This is why
the suppression of losses is achieved within a relatively broad frequency
range and the results are not particularly sensitive to the number of layers
in the periodic structure. By contrast, the possibility of a significant
size reduction discussed in the next subsection is associated with the
transmission band edge resonance, which is usually characterized by a
relatively narrow bandwidth and strong dependence on the number of layers in
the periodic stack.

\subsection{Resonant absorption suppression}

Under what circumstances can we not only suppress the absorption but also
have the size of the periodic composite structure much smaller than that of
the uniform (magnetic) slab with similar performance? When considering this
question we should keep in mind that within the framework of the photonic
crystal approach the characteristic length $L$ of the periodic array is
always comparable to that of the electromagnetic wavelength in the medium.
Therefore, for a given frequency range and for a given set of the
constitutive materials, we cannot significantly change the structural period 
$L$. Nor can we substantially reduce the number $N$\textit{\ }of unit cells
without loosing all the effects of coherent interference. For instance, in
our numerical example we have chosen $N=8$, and there is very little room
for further reduction in size. All we can achieve by adjusting the
configuration of the periodic array comprising as few as several periods is
to suppress the losses and/or to enhance the useful functionality, such as
Faraday rotation. The real question is: what is the thickness $D_{U}$ of the
uniform slab with the useful functionality comparable to that of the
optimized layered structure? Indeed, if such a uniform slab turns out to be
much thicker than the layered structure, then we can claim that not only the
periodic array dramatically reduces the losses, but it also has much smaller
dimensions. The latter is only possible if the thickness $D_{U}$ of the
uniform slab with desired functionality is much greater than the
electromagnetic wavelength in the medium. Otherwise, all we can achieve by
introducing periodic inhomogeniety is a significant reduction of losses
which, by the way, has been our primary objective.

Let us turn back to the periodic stack in Fig. \ref{FSn8} with the
permittivity and permeability tensors of the $F$ and $A$ layers given in (%
\ref{eps_F, mu_F}) and (\ref{eps_A, mu_A}). In the numerical example (\ref%
{set 1}), we assume a relatively strong circular birefringence $\beta =4.0$
of the magnetic material. The resulting Faraday rotation in a uniform
magnetic slab is of the order of unity even if the magnetic slab thickness
is equal to just several electromagnetic wavelengths. Such a small thickness
is comparable to the thickness of a photonic structure (a periodic stack)
comprising as few as several unit cells $L$. Therefore, in the case (\ref%
{set 1}) of a strong circular birefringence of the magnetic material, the
periodic stack cannot have significantly smaller size in comparison to the
uniform magnetic slab producing similar nonreciprocal Faraday rotation. The
composite structure in this case is only useful for absorption reduction, as
we have demonstrated in the previous subsection.

A strong circular birefringence usually occurs in the vicinity of a
ferromagnetic resonance at microwave frequency range. If the frequency $%
\omega $ is much higher or much lower than that, the gyrotropic parameter $%
\beta $ is much smaller. For instance, at infrared and optical frequencies,
in order to produce Faraday rotation $\rho $ of the order of unity, the
thickness of the uniform magnetic slab should exceed the electromagnetic
wavelength by at least two or three orders of magnitude. By contrast, the
periodic magnetic stack incorporating the same magnetic material can be much
thinner (of the order of several electromagnetic wavelengths), while
producing Faraday rotation $\rho $ of the order of unity and dramatically
reducing the absorption. In other words, the composite structure with the
proper configuration can suppress losses \textbf{and} be much smaller than
the uniform sample, but only if the circular birefringence $\beta $ of the
magnetic material is relatively weak, specifically, if%
\begin{equation*}
\beta \ll \mu _{1}.
\end{equation*}%
To illustrate this point, let us consider the following set of numerical
values describing the uniform magnetic slab in Fig. \ref{FSnU} and the
periodic magnetic stack in Fig. \ref{FSn8}

\begin{equation}
\begin{array}{c}
d_{A}=d_{F}=0.5L,\ N=8,\ D_{S}=8L,\ D_{U}=40L, \\ 
\ \varepsilon _{1}=2.89,\ \mu _{1}=32.49,\ \varepsilon _{0}=26.01, \\ 
\alpha =0,\ \beta =0.1,\text{\ \ }\gamma _{e}=0.01,\text{\ }\gamma _{m}=0.%
\end{array}
\label{set 2}
\end{equation}%
Again, the $F$ layers of the periodic array are made of the same lossy
magnetic material as the uniform magnetic slab. The Bloch dispersion
relation of the respective periodic structure is shown in Fig. \ref%
{DR_a05_b01}. There are two major differences between the numerical values
in (\ref{set 1}) and (\ref{set 2}). In the latter case, the value $\beta $
of specific Faraday rotation of the magnetic material is $40$ times smaller.
To partially offset the weaker circular birefringence and still have the
circular birefringence $\rho $ of the order of unity, the thickness $D_{U}$
of the uniform magnetic slab is set to be much larger ($40L$). By contrast,
properly designed periodic array can have just several unit cells $L$ and
sill produce the desired nonreciprocal effect. All the above can be achieved
if the frequency is close enough to one of the transmission band edge
resonances, also known as slow wave Fabry-Perot resonances. Two of such
resonances are shown in Fig. \ref{tf_a05_b01_LC}. As we shall see below, in
the case (\ref{set 2}) we can suppress the losses and reduce the size at the
same time, but all this comes at the expense of the bandwidth. Specifically,
the bandwidth of the desired effect of strong Faraday rotation in a
combination with a significant absorption suppression is now limited by the
width of the respective transmission band edge resonance. By contrast, in
the case (\ref{set 1}) described in Fig. \ref{af_pt_b4}, the dramatic
reduction of absorption occurs within a broad frequency range spanning a
significant portion of the lowest photonic band.

A graphic illustration of the effect of resonance absorption suppression is
presented in Fig. \ref{af_pt_b01}. A sharp peak in the frequency dependence
of the Faraday rotation $\rho $ in Fig. \ref{af_pt_b01}(a) coincides with
the transmission band edge resonance seen in Fig. \ref{tf_a05_b01_LC}.
Observe that the dramatic enhancement of the Faraday rotation in Fig. \ref%
{af_pt_b01}(a) only occurs in the vicinity of the transmission band edge
resonance and is characterized by a relatively narrow bandwidth. Away from
the resonance, the stack performance is now mixed. This situation is typical
for different resonance and slow-wave phenomena, such as the frozen mode
regime \cite{PRB03,PRE05,PRE06,PRA07}. By contrast, in our previous example (%
\ref{set 1}), the absorption suppression occurred within a broad frequency
range and it was not related to any resonance.

\section{Conclusion}

\begin{figure}[tbph]
\scalebox{0.8}{\includegraphics[viewport=-30 0 500 500,clip]{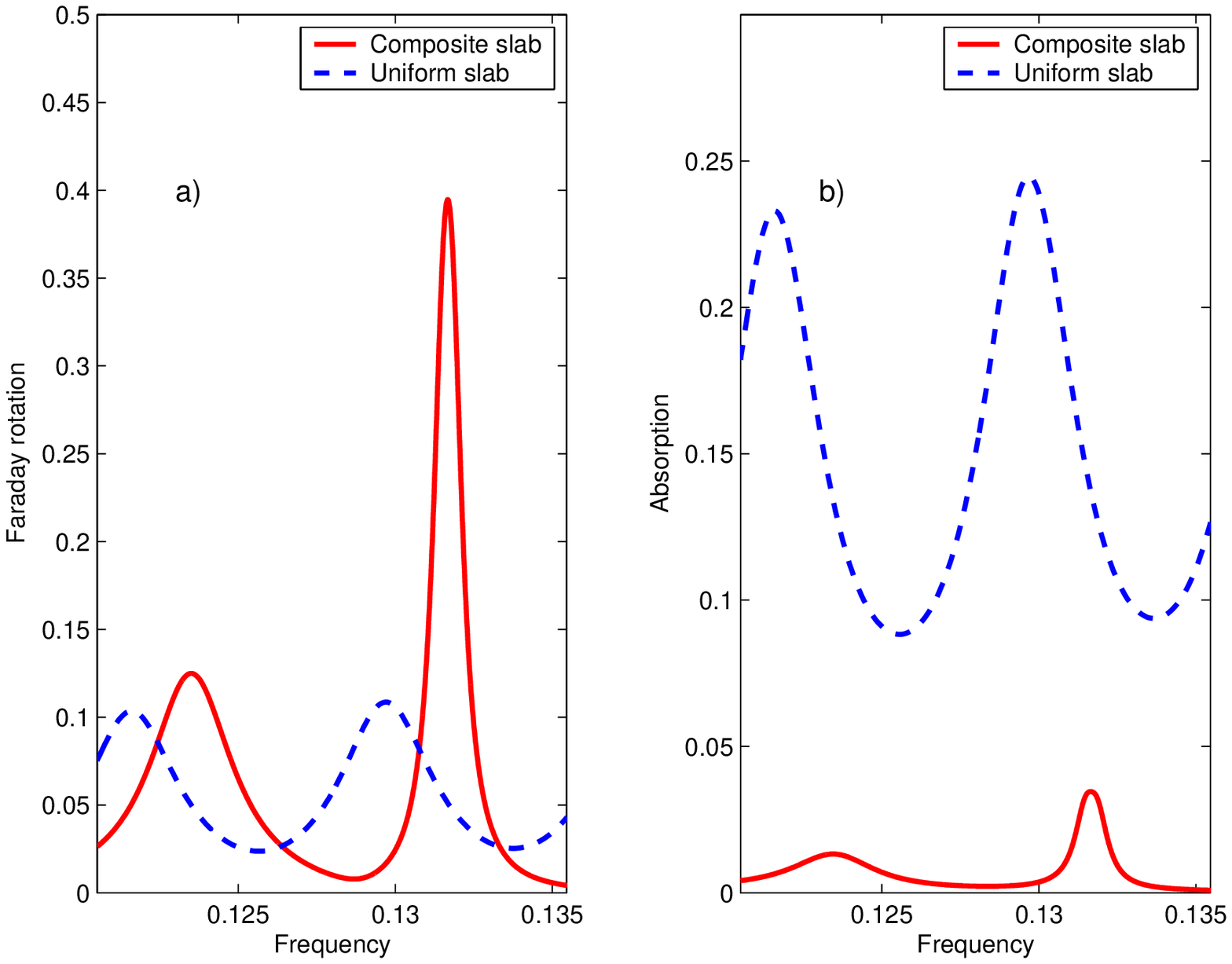}}
\caption{(Color online) Comparative transmission characteristics of the
periodic magnetic stack versus the uniform magnetic slab. The physical
parameters of both uniform and non-uniform samples are specified in (\protect
\ref{set 2}): (a) The Faraday rotation $\left\vert \protect\rho \right\vert $
versus frequency $\protect\omega $ of the uniform magnetic slab (the dashed
curve) and the periodic layered structure (the solid curve). (b) The
respective absorption $a$ versus frequency. The graphs show that at the
frequency of transmission band edge resonance, the composite structure
produces much stronger Faraday rotation than that of the uniform slab, while
greatly reducing the absorption. In addition, the periodic stack thickness
is 5 times smaller than that of the uniform magnetic slab.}
\label{af_pt_b01}
\end{figure}

We have shown that periodic composite structures, such as photonic crystals,
can be used to dramatically suppress the energy dissipation associated with
the presence of lossy component. In most cases, the useful functionality of
the lossy component can be preserved or even enhanced by the presence of
spatially periodic inhomogeniety. In our first example described in Fig. \ref%
{af_pt_b4}, the absorption suppression by the periodic structure is not a
resonance effect. As a consequence, it occurs in a relatively broad
frequency range and it is not particularly sensitive to the size and shape
of the photonic crystal. The optimal parameters of the periodic structure
depend on: (i) the useful functionality of the lossy material, such as
Faraday rotation, nonlinearity, etc., (ii) the physical nature of
absorption, and (iii) the frequency range of interest. In addition to a
significant reduction of losses, the properly designed periodic array can
also have much smaller dimensions, compared to the uniform sample with
similar functionality. But the size reduction is possible only in special
cases and it comes at the expense of the bandwidth, as illustrated in Fig. %
\ref{af_pt_b01}.

\textbf{Acknowledgments:} Effort of A. Figotin and I. Vitebskiy is sponsored
by the Air Force Office of Scientific Research, Air Force Materials Command,
USAF, under grant number FA9550-04-1-0359.

\bigskip\pagebreak

\bigskip\pagebreak

\begin{figure}[tbph]
\scalebox{0.8}{\includegraphics[viewport=-30 0 400 450,clip]{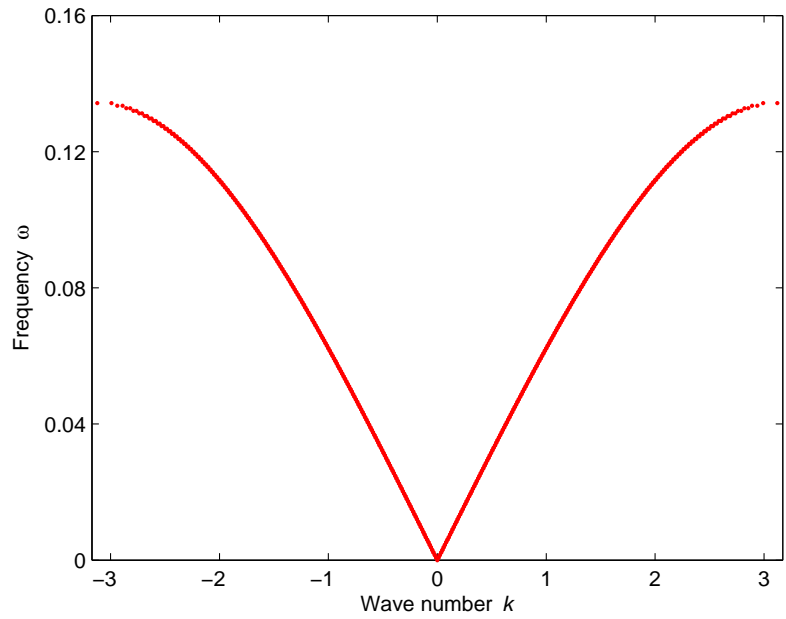}}
\caption{(Color online) The lowest band of the Bloch $k-\protect\omega $
diagram of the periodic layered structure in Fig. \protect\ref{FSn8} with
the physical parameters specified in (\protect\ref{set 2}). Due to the
relatively weak circular birefringence in magnetic layers, the split of the
spectral branches is not visible, as opposed to the case (\protect\ref{set 1}%
) shown in Fig. \protect\ref{DR_a08_b4}.}
\label{DR_a05_b01}
\end{figure}

\begin{figure}[tbph]
\scalebox{0.8}{\includegraphics[viewport=-30 0 500
400,clip]{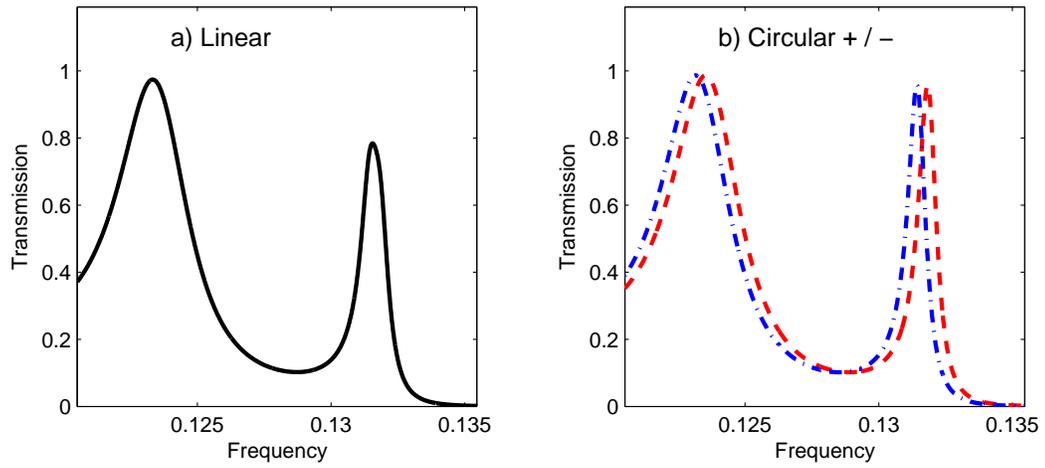}}
\caption{(Color online) Transmission $t$ versus frequency $\protect\omega $
of the periodic magnetic stack in Fig. \protect\ref{FSn8} with physical
parameters specified in (\protect\ref{set 2}): a) the incident wave
polarization is linear. b) the incident wave polarization is circular. Two
transmission maxima on the right correspond to two Fabry-Perot resonances
for each of the two senses of circular polarizations. At frequencies
corresponding to the curve intersection, the transmitted wave polarization
is strictly linear (a pure Faraday rotation). At all other frequencies,
there is a small ellipticity.}
\label{tf_a05_b01_LC}
\end{figure}

\end{document}